 \documentclass[final,5p,times,twocolumn,authoryear]{elsarticle}

\usepackage{amssymb}
\usepackage{lipsum}
\usepackage{url}
\usepackage{xcolor}
\usepackage{etoolbox}
\usepackage{aas_macros}
\makeatletter
\patchcmd{\@verbatim}
  {\verbatim@font}
  {\verbatim@font\small}
  {}{}
\makeatother
\newcommand{\merc}{\texttt{Mercury}}
\newcommand{\marxes}{\texttt{Mercury-Ar$\chi$es}}
\newcommand{\mopal}{\texttt{Mercury-Opal}}

\journal{Astronomy $\&$ Computing}

\begin{document}

\begin{frontmatter}



\title{Mercury-Opal: the GPU-accelerated version of the n-body code for planet formation Mercury-Ar$\chi$es}


\author[1,2]{Paolo Matteo Simonetti}
\affiliation[1]{organization={INAF - Osservatorio Astronomico di Trieste},%Department and Organization
            addressline={Via G.B. Tiepolo 11}, 
            city={Trieste},
            postcode={34143}, 
            %state={},
            country={Italy}}
\affiliation[2]{organization={ICSC National Centre for HPC, Big Data and Quantum Computing},%Department and Organization
            addressline={Via Magnanelli 2}, 
            city={Casalecchio di Reno},
            postcode={40033}, 
            %state={},
            country={Italy}}

\author[3,2]{Diego Turrini}
\affiliation[3]{organization={INAF - Osservatorio Astrofisico di Torino},%Department and Organization
            addressline={Strada Osservatorio 20}, 
            city={Pino Torinese},
            postcode={10020}, 
            %state={},
            country={Italy}}

\author[4,2]{Romolo Politi}
\affiliation[4]{organization={INAF - Istituto di Astrofisica e Planetologia Spaziali},%Department and Organization
            addressline={Via Fosso del Cavaliere 100}, 
            city={Rome},
            postcode={00133}, 
            %state={},
            country={Italy}}

\author[4]{Scig\'e J. Liu}

\author[4]{Sergio Fonte}

\author[3,2]{Danae Polychroni}

\author[1,2,5]{Stavro Lambrov Ivanovski}
\affiliation[5]{organization={University of Trieste},%Department and Organization
            addressline={Piazzale Europa, 1}, 
            city={Trieste},
            postcode={34127}, 
            %state={},
            country={Italy}}            
            
\begin{abstract}
Large n-body simulations with fully interacting objects represent the next frontier in computational planetary formation studies. In this paper, we present \mopal, the GPU-accelerated version of the n-body planet formation code \marxes. % \marxes\, is a branched-out version of \merc, a general purpose n-body integrator, and includes the gravitational and viscous interaction between planetesimals and the protoplanetary disk, plus the accretion processes. 
The porting to GPU computing has been performed through OpenACC to ensure cross-platform support and minimize the code restructuring efforts %and an high level of serviceability even by researchers not specifically trained in GPGPU computing 
while retaining most of the performance increase expected from GPU computing. We tested \mopal\, against its parent code \marxes\, under conditions that put GPU computing at disadvantage and nevertheless show how the GPU-based execution provides advantages with respect to CPU-serial execution even for limited computational loads.
%\mopal\, execution time increases only by a factor of 1.5 increasing the amount of fully interacting bodies from 1 to 1000, while the serial version execution time increases by a factor of 190 over the same interval. When all factors are considered, the speedup obtained by \mopal\, is equal to 2.2 in absolute terms and 16 in relative terms for the 1000 body case. 
\end{abstract}




\end{frontmatter}

%\tableofcontents



\section{Introduction}
\label{introduction}

\merc\, \citep{chambers1999} is a general purpose n-body integrator widely used in the planetary and exoplanetary science communities. At the time of writing, \merc\, has been used in more than 1600 studies and has served as the basis for multiple specialized codes that expanded its modeling capabilities by introducing new physical processes
%, continually developed over the last two and a half decades to serve a wide variety of needs in planetary sciences <- Tecnicamente l'ultima versione ufficiale di Mercury è quella del 2001.
%Several versions of this code have been branched out during the years, 
such as spin-orbit interaction \citep[\texttt{SMERCURY,}][]{lissauer2012}, tides \citep[\texttt{Mercury-T,}][]{bolmont2015}, the interaction between planetary bodies and their native protoplanetary disk \citep[\marxes,][see also Turrini et al., this issue]{turrini2019,turrini2021} and the Yarkovsky-O'Keefe-Radzievskii-Paddack effect \citep[][]{fenucci2022}. The main limiting factor to the scientific use of \merc, shared by all n-body codes, is the number of planetary bodies that can be modeled, which is constrained by the computational efficiency of the adopted dynamical algorithm, the timescale of the problem to be modeled and, above all, the available computational resources.

The growth of computing power supplied by multi-cores architectures offered the possibility for increasingly detailed simulations and boosted the transition of n-body codes to parallel programming, e.g. \texttt{REBOUND} \citep[][]{rein2012}, \texttt{PKDGRAV3} \citep[][]{potter2017}, \marxes\,\citep{turrini2019,turrini2021}. As a result, while studies at the turn of the century could include either $\sim 10^2$ bodies integrated for $10^8$ years \citep[e.g.][]{chambers1998,chambers2001}, or $\sim 10^6$ bodies integrated for just $10^3$ years \citep[][]{richardson2000}, current calculations can integrate $10^5$ bodies for up to $10^7$ years \citep[][]{woo2021}.
GPU accelerators constitute the next iteration of this approach, as they increase both the computational density and energy efficiency \citep{portegies2020} while decreasing the hardware cost on a per-operation basis. Example of existing GPU-parallel n-body codes are \texttt{HiGPU} \citep[][]{capuzzo2013}, \texttt{GENGA} \citep[][]{grimm2014} and \texttt{GENGA II} \citep[][]{grimm2022}, and \texttt{GLISSE} \citep[][]{zhang2022}.

The effective use of GPUs requires careful data management to comply with the more stringent limits of the GPU's internal RAM and the limits in the data transfer bandwidth between CPU (\textit{host}) and GPU (\textit{device}). The transfer of data and instructions between host and device and the parallel execution of sections of codes on GPUs can be achieved in three main ways. The first one is via a low-level programming application interface (API) like OpenCL\footnote{\url{https://www.khronos.org/opencl/}}, CUDA\footnote{\url{https://docs.nvidia.com/cuda/}}, HIP\footnote{\url{https://rocm.docs.amd.com/projects/HIP/en/latest/}} or SYCL\footnote{\url{https://www.khronos.org/sycl/}}. The second one is via high-level compiler directives such as OpenACC\footnote{\url{https://www.openacc.org/}} and OpenMP\footnote{\url{https://www.openmp.org/}}. The third one is via specific abstraction libraries, for example Kokkos\footnote{\url{https://kokkos.org/kokkos-core-wiki/}}, Raja\footnote{\url{https://github.com/LLNL/RAJA}}, Thrust\footnote{\url{https://developer.nvidia.com/thrust}} or the C++ Parallel Standard Template Library (PSTL)\footnote{\url{https://www.intel.com/content/www/us/en/developer/articles/guide/get-started-with-parallel-stl.html}}. Each approach differs in what hardware components they target and how much finer control they allow to or require from users, its portability and how much code restructuring they mandate in general.
%The transfer of data between host and device and the parallel execution of sections of codes on GPUs can be achieved using low-level programming application programming interfaces (APIs) like OpenCL and CUDA, or through the directive-based programming model OpenACC. 
%
%To overcome this barrier and increase the accessibility to GPU acceleration of the research community -which for the most part is not specialized in high-performance computing techniques-, OpenACC was introduced in November 2011\footnote{\url{https://www.openacc.org/sites/default/files/inline-files/OpenACC_1_0_specification.pdf}}. OpenACC is an Application Programming Interface (API) which 
%OpenACC allows for managing the parallelization on GPU at compiler level, while requiring the user only to insert high-level compiler directives (also called decorators in Fortran) in the code. Two advantages of OpenACC are the minimization of deep code restructuring and the possibility of creating GPU-capable codes building on top of the bottleneck and optimization analysis previously performed to parallelize codes on CPUs with OpenMP. 
In particular, OpenACC offers two advantages that we decided to leverage on for this work. The first is the minimization of deep code restructuring, associated with the use of compiler directives. The second one is its better performances and slightly more precise control granted to the user with respect to the competing OpenMP solution.

In this paper, we present the porting to GPU computing of \marxes\, \citep[][see also Turrini et al., this issue]{turrini2019,turrini2021}, a parallel n-body planet formation code built on top of \merc, %a \merc\, derivative developed specifically to study the interaction between planetesimals and a time evolving protoplanetary disk. 
which has been carried on under the OpenACC 3.3 specification\footnote{\url{https://www.openacc.org/sites/default/files/inline-images/Specification/OpenACC-3.3-final.pdf}}. This GPU-capable version of \marxes, which is called \mopal, has been developed in the context of the \textit{Origins of Primordial Atmospheres for arieL} (OPAL, Polychroni et al., this issue) Key Science Project of the Italian National Centre for HPC, Big Data and Quantum Computing (ICSC) with the goal of providing the community working on the preparation of the ESA Ariel mission \citep[][]{tinetti2018,turrini2018,turrini2022} with %a fast, interoperable and easy to service 
a physically realistic and efficient n-body integrator for formation-to-observation simulation campaigns.

The structure of this paper is as follows: in Sect.~\ref{sec:model} we present the salient features of the parent code and its profiling; in Sect.~\ref{sec:mopal} we delineate the parallelization strategy and show pseudocode blocks of the main loops ported on GPU; in Sect.~\ref{sec:results} we show the results in terms of performance and accuracy; in Sect.~\ref{sec:discussion} we briefly discuss our choices in terms of parallelization framework; finally, in Sect.~\ref{sec:conclusion} we identify the main areas of improvement and our future plans.

\section{The \marxes\, code}
\label{sec:model}

\begin{figure*}
\centering 
\includegraphics[width=0.98\textwidth]{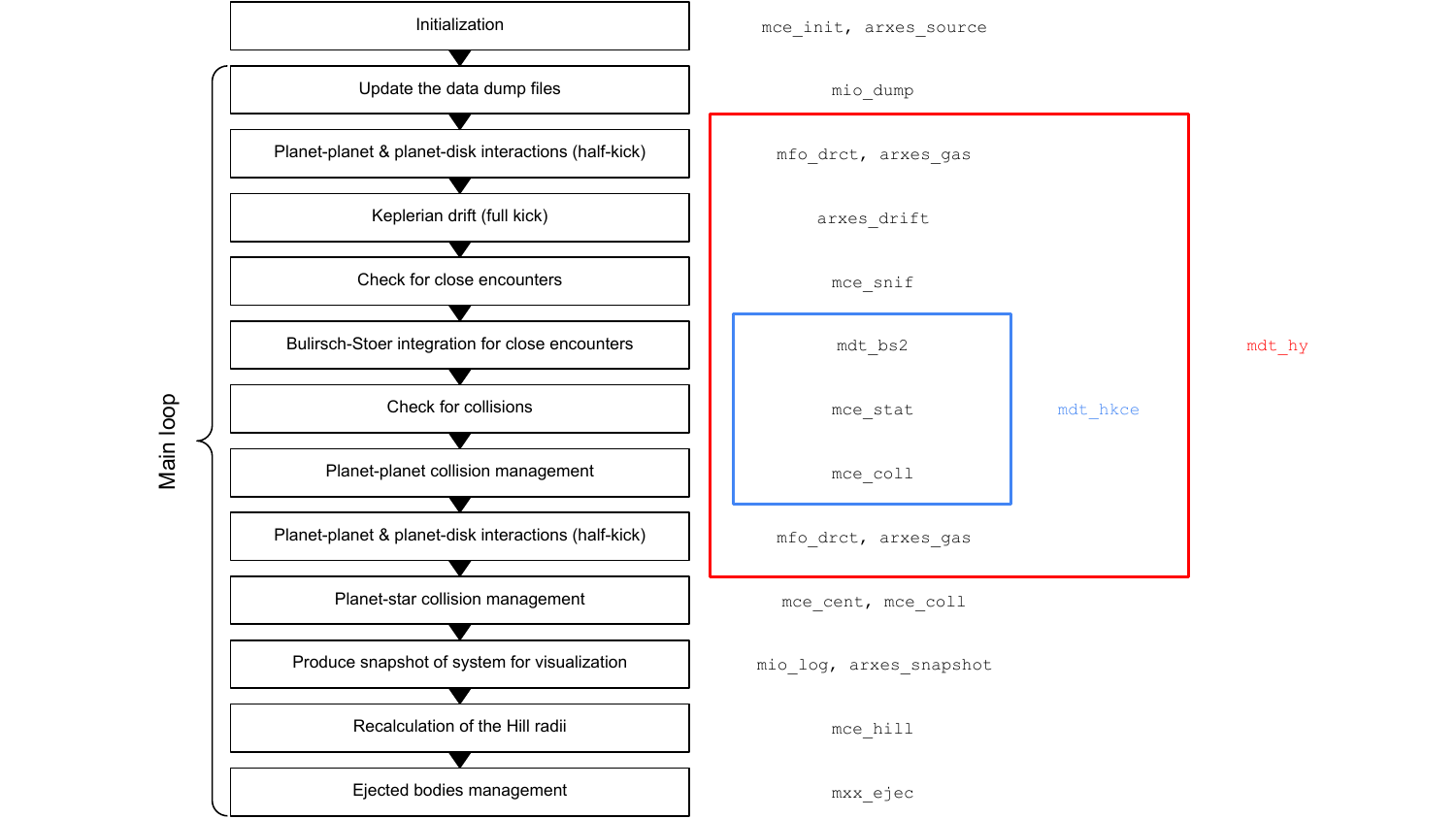}	
\caption{The general structure of the \marxes\, code. The left hand column represent the logical structure of the code, while the right hand column reports the names of the main subroutines involved in the computation. Higher level subroutines are reported in blue and red.} 
\label{fig:main_loop}
\end{figure*}

\subsection{Main features of the model}
\label{ss:marxes}

The starting point of our parallelization effort is \marxes\,\citep[][Turrini et al., this issue]{turrini2019,turrini2021}. This version of \merc\, has been designed to study the dynamical evolution over Myr-long timescales of forming planetary systems embedded within their native protoplanetary disk.  \marxes\, is a multi-language code that combines Fortran, specifically the FORTRAN77 standard of \merc\, and the Fortran95 of the Ar$\chi$es library, and C in the form the C99 library \texttt{WHFAST} \citep[][]{rein2015}. The most important advancements, in terms of the physical recipes, introduced by \marxes\,with respect to \merc\,are:
\begin{itemize}
\item a faster and more accurate subroutine for the keplerian position and velocity propagation of the bodies in the gravity well of the parent star (called \textit{keplerian drift}), which takes advantage of the \texttt{WHFAST} library \citep[][]{rein2015};
\item the evolution in mass, physical radius and semimajor axis of forming and migrating planets;
\item the gravitational interaction between the protoplanetary disk and the planetesimals embedded in it;
\item the non-gravitational interaction exerted by the gas in the protoplanetary disk and the planetesimals through aerodynamic drag.
\end{itemize}
\marxes\, has also been parallelized on CPU via the inclusion of OpenMP directives. As such, \marxes\, can be compiled either for serial or intra-node CPU-parallel execution. For a complete description of the code and the underlying physical library, we direct the reader to Turrini et al., (this issue) and \citet{turrini2019,turrini2021}.

While all planetary bodies in \marxes\, are sensitive to both gravitational and non-gravitational forces, only the massive ones, labeled as \textit{big bodies} in the software, are also gravity sources. Examples of such big bodies in a typical planet formation simulation are the Mars-sized planetary embryos that will collisionally grow to become planets. In addition to big bodies, \marxes\, allows to model the dynamical evolution of \textit{small bodies}, massless particles representing the planetesimal disk surrounding the forming planets. Small bodies are influenced by the gravity of the forming planets as well as by their interactions with the protoplanetary disk. The computational load is therefore driven by the largest term between the computation of the gravitational forces, which scales as $\mathcal{O}$(($N_{small}+N_{big}) \times N_{big})$ where $N_{big}$ is the number of massive bodies and $N_{small}$ that of massless particles, and the computation of the interactions between planetary bodies and the protoplanetary disk, which scales as $\mathcal{O}$($N_{small}+N_{big}$) but with higher computational density per body than the computation of the gravitational forces (see Turrini et al., this issue).
%This distinction allows to substantially reduce the integration time, since the number of body-body interactions to take into account in the \texttt{opal\_force} routine is $\propto N_{small} \times N_{big}$, where $N_{big} \sim 10$ is the number of \textit{big bodies}, while $N_{small} \sim 10^6$ is the number of planetesimals. 

%This choice is computationally advantageous but substantially reduces the robustness of the model predictions in the early phases of the protoplanetary disk evolution, since in a realistic scenario, at the beginning of the simulation most of the mass is in the planetesimals. Therefore, the need to overcome the large increase in computational resources needed to manage a fully interacting set of at least $\sim10^5$ bodies constitutes the main scientific motivation behind the transition to the GPU-accelerated framework presented in this paper.

In modern planet formation simulations, $N_{big}$ generally ranges between 1 and 100, while $N_{small}$ is of the order of $\sim 10^4 - 10^6$ \citep[e.g.][]{turrini2023,polychroni2025}. This choice is computationally advantageous but impacts the insight that can be obtained by models as the back reaction of the planetesimal disk on the forming planets cannot be directly modeled.
Modeling large numbers of fully interacting bodies is challenging using CPU-based parallelism, as it requires either extremely long integration times or large-scale computational clusters and the use of multi-node parallelization with MPI. The need to overcome this large increase in computational load needed to manage a fully interacting set of at least $\sim10^5$ bodies constitutes the main scientific motivation behind the transition to the GPU-accelerated framework presented in this paper.

%the robustness of the model predictions in the early phases of the protoplanetary disk evolution, since in a realistic scenario, at the beginning of the simulation most of the mass is in the planetesimals. Therefore, the need to overcome the large increase in computational resources needed to manage a fully interacting set of at least $\sim10^5$ bodies constitutes the main scientific motivation behind the transition to the GPU-accelerated framework presented in this paper.

\subsubsection{Structure of the code}
\label{sss:main_loop}

The main computation loop of \marxes\, is contained within the \texttt{mal\_hcon} routine called in the \texttt{main}. With the exception of the initial and final I/O operations, the final data dump at the very end of the simulation, all calculations are performed within the \texttt{mal\_hcon} call. \merc, being a general-purpose code, allows the user to choose among several different integration schemes with varying degrees of numerical robustness and speed. Because of the concurrent need for both energy conservation and handling of close encounters and collisions of a planetary formation simulation, \marxes\, always uses the second-order mixed-variable symplectic integrator included in \merc. The specific structure of this algorithm is schematically represented in Figure \ref{fig:main_loop}, together with the names of the main involved subroutines and their specific role in the computation. Aside for those explicitly mentioned, the code makes use of smaller support routines for:
\begin{enumerate}
\item the conversion of the bodies' coordinates from heliocentric to \textit{democratic heliocentric}\footnote{This entails the use of barycentric velocities and heliocentric positions, as needed to split the Hamiltonian of the system into the three separate components of the Keplerian motion (or drift) under the effect of the stellar gravity, the planet-planet interaction and the stellar motion \citep[][]{chambers1999}.} (\texttt{mco\_h2dh}) and back (\texttt{mce\_dh2h}), from heliocentric to barycentric (\texttt{mce\_h2b}), and from 3D velocity and positions to the osculating orbital elements (\texttt{mxx\_x2el}, called by \texttt{arxes\_snapshot});
\item the conversion of the time from the Julian Day to the Gregorian calendar date (\texttt{mio\_jd2y});
\item the copy creation of position and velocity arrays (\texttt{mco\_iden});
\item the calculation of physical and critical radii (\texttt{mce\_init}) of the bodies;
\item the calculation of the boxes around each body to identify the close encounters (\texttt{mce\_box}) and the encounters minima (\texttt{mce\_min});
\item the elimination of lost bodies from the mass, position and velocity arrays (\texttt{mxx\_elim});
\item the total energy and momentum of the system (\texttt{mxx\_en});
\end{enumerate}    
These support routines are called several times during the execution. For example, those at (4) and (6) are called whenever the number or size of the bodies in the system changes, i.e. after collisions and ejections. The \texttt{mce\_hill} routine is called at each timestep since the Hill radius\footnote{The 3D surface within which the gravitational force exterted by the body dominates over that of the central star.} of each body depends on the star-planet separation, but also within the \texttt{mce\_init} as part of the critical radius\footnote{The critical radius, which is a multiple of the Hill radius, is used by the code to determine if it is safe or not to advance the interaction part of the Hamiltonian using the hybrid symplectic integrator, which conserves the total energy of the system only if the hierarchy within the system is conserved. If a body enters the Hill Sphere of another body, this condition is violated and \marxes\, must switch to the slower Bulirsch-Stoer integration scheme to advance the interaction and Keplerian Hamiltonians of the bodies involved in the close encounter \citep[][]{chambers1999}.} recalculation after impacts of bodies with the central star and expulsions form the system.

\subsubsection{Profiling of the code}
\label{sss:marxes_prof}

\begin{figure}
\centering 
\includegraphics[width=0.48\textwidth]{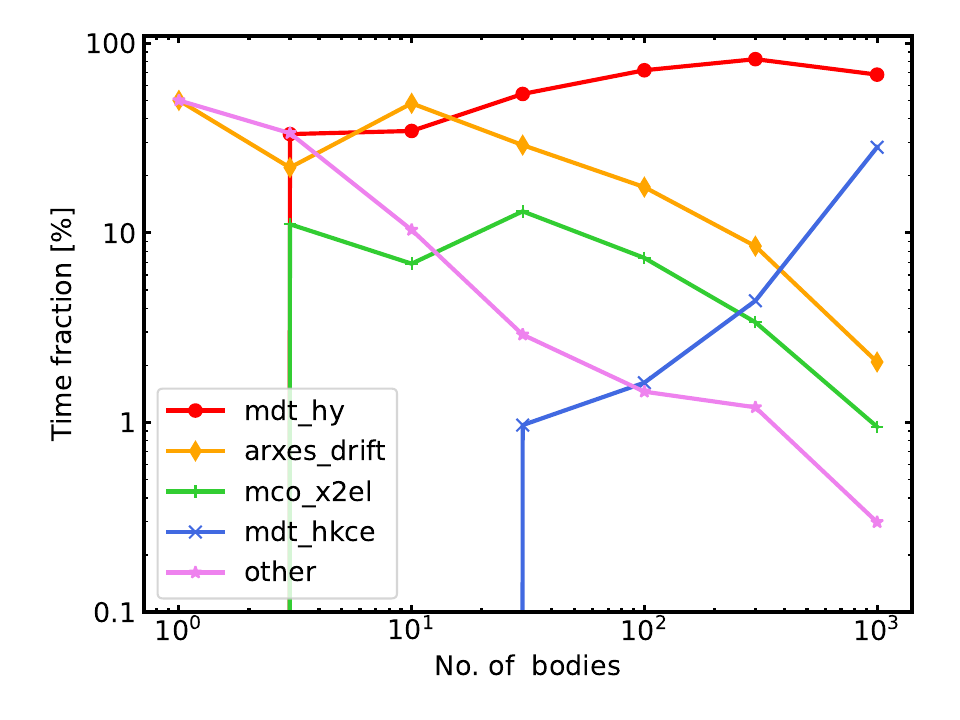}	
\caption{The \marxes\, profile as obtained by \texttt{gprof} for runs with different amount of bodies over an integration time of 1000 yr. \texttt{mdt\_hy} (in red) refers to all the operations in that subroutine with the exclusion of \texttt{arxes\_drift} and \texttt{mdt\_hkce}, which are plotted separately (in orange and blue, respectively); \texttt{mco\_x2el} (in green) is called by \texttt{arxes\_snapshot}; \textit{other} (in purple) account for the sum of all the other subroutines.} 
\label{fig:time_profiling}
\end{figure}

The first step in the GPU-porting process was to profile the serially executed \marxes\, code. We tested seven different configurations with an increasing number of fully gravitating bodies ($N_{big}$=1, 3, 10, 30, 100, 300 and 1000), integrated for 1000 yr. In all cases, $N_{small}=0$. Under this choice, computational complexity depends only on one variable, $N_{big}$, which from now on will be referred to simply as $N$. The architecture of the simulated planetary systems include a Solar-mass central star, orbited by four already-formed massive planets with masses in the [0.45, 2.2] Jupiter masses range and semi-major axes in the [0.31, 3.97] AU range, and a variable number of smaller bodies of $10^{-2}$ Earth masses with randomly distributed semi-major axes in the [1, 10] AU range. When $N<4$ we simply exclude one or more of the most massive objects. Each configuration was run five times and the shown results, reported in Figure ~\ref{fig:time_profiling}, are the arithmetic means of the results. The time profiling has been obtained by using \texttt{gprof}\footnote{\url{https://sourceware.org/binutils/docs-2.38/}} v.2.38.

%First of all, both the total wall time and the fractional time distributions among the single subroutines are very similar irrespective of the presence or not of the non-gravitational forces. This happens because the gas drag is calculated anyway and simply not added to the total acceleration. As such, we will discuss the configurations without them and generalize our conclusions. Apart for the data dumps (red curve), which dominates the time budget even at $N=1000$,

%The results are reported in Figure ~\ref{fig:time_profiling}. 
As it is possible to see, most of the computation time for $N\ge10$ is spent in the \texttt{mdt\_hy} subroutine which in the plot is represented by the sum of the red, yellow and blue curves. If we analyze more closely this subroutine, we notice that the weight of the specific components changes with $N$. The calculation of the keplerian drift is the largest contributor when $N \lesssim 10$, where it takes up to about 50\% of the time, while for $N \gtrsim 30$ the calculation of the forces overtakes the role of the most computationally expensive part of the code, reaching a time fraction of $\sim68\%$ at $N=1000$. At $N \sim 300$ the close encounter management, performed in \texttt{mdt\_hkce}, takes up 4\% of the total time but above this threshold it grows rapidly, reaching 28\% at $N \sim 1000$. The growth of both \texttt{mdt\_hy} and \texttt{mdt\_hkce} is not surprising: when all the bodies are fully interacting with each other, the number of body-body interactions increase as $\propto N^2$, while the weight of other operations such as coordinate conversion (e.g., \texttt{mco\_x2el}, green curve) or I/O scale linearly with $N$. The larger slope of the \texttt{mdt\_hkce} curve with respect of the \texttt{mdt\_hy} curve is probably related to the more complex structure of the close encounter evaluation that is associated with multiple if branches and larger computational load per particle than other subroutines. %, that require a sizeable number of if branches. 
%On the other hand
Conversely, in the interaction and the Hamiltonian calculations the if branches are both less frequent and simpler, with several of them being controlled by input parameters that remain constant throughout the simulation.

%At variance with 
While close encounters are continuously evaluated during the simulations, proper collisions and expulsions are rare. As such, planet-planet and planet-star collision and ejection management routines are called only sparsely during execution. In our simulation setup, ejections are checked for only once every 200 years by searching for bodies whose separation from the central star exceeds 400 AU. In the configuration tested here, this operation is performed only five times and its weight grows proportionally to $N$. As an example, we can consider the specific case with $N=1000$: in that configuration, the \texttt{mce\_coll} routine, which is used both for planet-planet collisions within \texttt{mdt\_hkce} and for planet-star collision outside \texttt{mdt\_hy}, is called a total of $\sim10^2$ times while \texttt{mdt\_hy} is called $\sim10^5$ times. Since both the direct computation of the forces and the number of collisions are proportional to $N^2$, the proportion of time spent by the code managing collisions is not expected to increase even at larger $N$. The \textit{others} (purple) curve in Figure ~\ref{fig:time_profiling}) is the sum of the time fraction spent in all the other routines. While this value is in fact very large ($\sim$50\%) for the 1-body case, it rapidly decreases to 0.3\% at $N=1000$. Within this category, at small ($\lesssim 100$) $N$ the largest contributors are the coordinate conversion routines \texttt{mco\_h2dh} and \texttt{mco\_dh2h}, while at large ($\gtrsim 300$) $N$ this role is overtaken by the routine calculating the close encounter "boxes" around each planet \texttt{mce\_box} and the routine checking for collisions with the central body \texttt{mce\_cent}. The explanation likely resides in the fact that both of them contain a relatively high number of if branches, which increases the computational weight per body. Since all the operations involved in the execution of these routines scale as $N$, their relative contribution to the total execution time decreases when $N$ increase. Additional profiling discussion is provided in Turrini et al., this issue.

From the trends shown in Figure ~\ref{fig:time_profiling} it is easy to see that the computational gain from GPU-porting, especially when the number of interacting bodies is very large ($N \gtrsim 10^5$), will be driven by \texttt{mdt\_hy}. We therefore focused our efforts on this part of the code.
%The trend shown in Figure ~\ref{fig:time_profiling} are expected to continue, since they can be easily explained by the algorithmical complexity of each segment of the code. As a result, it is clear that most of the computational speed increase that can be obtained from parallelization, especially when the number of interacting bodies is very large ($N \gtrsim 10^5$), resides in \texttt{mdt\_hy}. Therefore, we elected to target that part of the code for GPU porting.

\section{The GPU porting of \marxes: \mopal}
\label{sec:mopal}

%After having identified the main time sinks, we proceeded to parallelize them. Luckily enough, 
Our analysis, as well as the one performed to parallelize \marxes\, on CPU (see Turrini et al., this issue), confirms that most of the calculations can be easily ported on a GPU since (i) the same set of instructions can be be applied to all bodies, (ii) there is a low number of \textit{if branches} within the \textit{do loops} and (iii) there is a low number of element-element dependencies within the \textit{do loops}. The only exception is the Bulirsch-Stoer integration, for which a (generally small) subset of all the bodies undergoes lengthier calculations involving a complex structure of nested, interdependent loops populated by if branches. While the computational density per body of \texttt{mdt\_hkce} is higher than that of \texttt{mdt\_hy}, the possible gain achievable through parallelization is severely limited by the low number of bodies that are processed at each timestep within \texttt{mdt\_hkce}. This is a direct consequence of the fact that, under most realistic scenarios, only a very small ($\lesssim 1$\%) number of bodies undergo close encounters at any given timestep. Moreover, each close encounter requires a different number of iterations to reach the required precision, which translate in an intrinsically unbalanced load. On the other hand, leaving the \texttt{mdt\_hkce} execution on the host requires only a limited volume of data to be moved to the CPU and back. As such, based on these considerations, in \mopal\,we decided to keep the \texttt{mdt\_hkce} subroutine on the host. Further discussion on the scaling of computational complexity with $N$ is provided in Sect.~\ref{sec:discussion}.
To proceed with the GPU-porting, we were forced to unpack some of the largest routines in order to avoid stack size problems. Stack size in the OpenACC framework is limited to 512 kB. While this is more than sufficient to store the algorithm, it cannot host some of the large local temporary arrays created by the routines themselves. As such, in \mopal, the instructions previously in the \texttt{mfo\_drct}, \texttt{arxes\_gas} and \texttt{mce\_snif} have been moved directly to \texttt{mdt\_hy}.

Another change we introduced in the code is the conversion of the \texttt{WHFAST} library integrated into \marxes\, from its original C99 implementation \citep{rein2015} into Fortran. This conversion has been done to address a recurring compilation error that prevented the correct operation of the Fortran-C interface module. Alongside solving the issue and simplifying the maintenance of the code, this conversion had the side benefit of slightly improving the overall efficiency of the code in the speed test of the serial execution of $\mopal$ when comparing the Fortran-C99 and the Fortran-only versions of the symplectic integrator, both compiled with the Intel OneAPI HPC toolkit. %\texttt{ifx/icx} Intel package. 
Our test revealed a small decrease in the total computation time, which is proportional to the computational weight of the \texttt{arxes\_drift} routine, using the Fortran-only version. % of the code which, despite being small, is still welcomed.

\subsection{Loop parallelization on GPU}
\label{ss:mopal_loops}

%Loop parallelization has been performed by means of the \textit{!\$acc parallel loop} instruction. For each loop we defined a dedicated parallel region following the guidelines of the \textit{OpenACC Programming and Best Practices Guide}\footnote{\url{https://www.openacc.org/sites/default/files/inline-files/OpenACC_Programming_Guide_0_0.pdf}}. %We also constructed a version of \mopal\, where only two, large parallel regions encapsulating respectively all the do loops before and after the \texttt{mdt\_hkce} call were defined. While there was a marginal increase in the computational speed (few seconds for the $N=1000$ case, whose average execution time is 338 s), we incurred in the risk of performing some of the calculations in the wrong order, which manifested itself in NaN outputs at random points during the execution.
There are five large code blocks that have to be parallelized and ported to the GPU, represented in Fig.~\ref{fig:main_loop} by the logical blocks within \texttt{mdt\_hy} but outside \texttt{mdt\_hkce}: the two half-timestep gravitational (\texttt{mfo\_drct}) and non-gravitational (\texttt{arxes\_gas}) forces calculation, the full timestep keplerian drift (\texttt{arxes\_drift}), the check for collisions (\texttt{mce\_snif}) and the evaluation of the interaction and stellar Hamiltonian at each half-step. In the following pseudocode blocks the reader will notice that some loops start from 2, rather than 1. The reason is that the first element of the mass, radius, position, velocity and acceleration arrays is reserved for the central body of the system, namely the star, and some of the calculations (e.g. the disk gap width) are performed only on planetary objects.

Each of these blocks have a different level of complexity. As such, their porting on GPU entailed slightly different strategies. Below we report four pseudocode samples from the \texttt{mdt\_hy} subroutine, showing in practical terms how we proceeded for this work. The first one represent the forces calculation:
\begin{verbatim}
!$acc parallel
<scalar variables initialization>
! LOOP 1
!$acc loop
        do i=1,nbod
          <vector variables initialization>
        end do
!$acc end loop
! LOOP 2
!$acc loop independent
        do j = 2, nbod
          asx=0.d0
          asy=0.d0
          asz=0.d0
!$acc loop reduction(+:asx,asy,asz) independent
          do i = 2, nbig
            <asx, asy, asz computation>
          end do
!$acc end loop
          a(1,j)=asx
          a(2,j)=asy
          a(3,j)=asz
        end do  
!$acc end loop
! LOOP 3
!$acc loop
          do i = 2, nbig
            <computation of disk gap widths>
          end do
!$acc end loop
! LOOP 4
!$acc loop
          do j = 1, nbod
!$acc loop seq
              do i = 2, nbig
                <checks if body j crosses the disk 
                gap produced by body i>
              end do
!$acc end loop
          end do
!$acc end loop
! LOOP 5
!$acc loop private(<gas drag and gravity calculation
                   aux vars>)
          do j = 1, nbod
            <aerodynamic drag computation>
            <gas gravity computation>
            a(1,j) = a(1,j) + drag + gas gravity on x
            a(2,j) = a(2,j) + drag + gas gravity on y
            a(3,j) = a(3,j) + drag + gas gravity on z
          end do
!$acc end loop
!$acc end parallel
\end{verbatim}
where \texttt{nbod} is $N$, \texttt{nbig} is $N_{big}$ and \texttt{a(3,nbod)} are the accelerations along the three axes of each body. As it is possible to see, this calculation consists of five loops, two of which (loop 2 and loop 4) have a computational complexity of $\mathcal{O}(N^2)$ and one (loop 5) scales as $N$ but is computationally dense. In loop 2 we had to include the \texttt{independent} keyword to ensure proper parallelization of the computation, while in loop 5 we included a list of scalar private variables explicitly, even though the compiler is expected to automatically define them as such, to improve code readability (see \textit{OpenACC Programming and Best Practices Guide}\footnote{\url{https://www.openacc.org/sites/default/files/inline-files/OpenACC_Programming_Guide_0_0.pdf}}).

The second block that we show concerns the calculation of the interaction and stellar Hamiltonian of the system:
\begin{verbatim}
!$acc parallel
! LOOP 1
!$acc loop
      do j = 2, nbod
        <velocity calculation>
      end do
!$acc end loop
! LOOP 2
!$acc loop reduction(+:mvsum(:3))
      do j = 2, nbod
        <summation of Hamiltonian terms>
      end do
!$acc end loop
! LOOP 3 
!$acc loop
      do j = 2, nbod
        <positions calculation>
      end do
!$acc end loop
! LOOP 4
!$acc loop
      do j = 1, nbod
        <store a copy of positions and velocities>
      end do
!$acc end loop      
!$acc end parallel
\end{verbatim}
These calculations are computationally light, since their complexity scales as $\mathcal{O}(N)$ and the number of operations per body is low. Loop 2 entails a $\texttt{reduction}$ operation, while the other loops are easily parallelized by the compiler without additional directives.

The third block consists of a large routine, \texttt{arxes\_drift}, which is executed sequentially on each CUDA core. As such, the code in \texttt{mdt\_hy} appears in this form:
\begin{verbatim}
<create a copy of x and v called x0 and v0>
!$acc parallel loop private(iflag) independent
      do j = 2, nbod
        iflag = 0
        call arxes_drift(m(1),x(1,j),x(2,j),
     %       x(3,j),v(1,j),v(2,j),
     %       v(3,j),h0,iflag)
      end do
\end{verbatim}
where \texttt{iflag} signals if the integration was successful or not, while the \texttt{arxes\_drift} routine was marked with the \texttt{!\$acc routine seq} directive. Copies of the positions and velocities before the keplerian propagation are created in case close encounters are detected by the \texttt{mce\_snif}. They will be used by \texttt{mdt\_hkce} to solve the close encounters using a Bulirsch-Stoer integrator.

%\textcolor{red}{Since the management of the close encounters requires the position and velocities of all the bodies both at the beginning and at the end of the timestep, described respectively by \texttt{x0(3,nbod)}, \texttt{v0(3,nbod)}, \texttt{x(3,nbod)} and \texttt{v(3,nbod)}, we asynchronously move the first two arrays from the device to the host while we calculate the latter two. The \texttt{!\$acc wait} directive is needed in order to avoid the output of NaN values as results, if the code proceeds before the transfer is completed.}

Finally, the fourth block that we ported on GPU is \texttt{mce\_snif}:
\begin{verbatim}
! LOOP 1
!$acc parallel loop private(<box calc aux vars>)
      do j = 2, nbod
        <finds the coordinates extremes
        of each body within this timestep>
      end do
!$acc end parallel loop
! LOOP 2
!$acc parallel loop
      do j = 2, nbig
        <adds rcrit to the coordinate extremes>
      end do
!$acc end parallel loop
! LOOP 3
!$acc parallel loop private(<encounter aux vars>)
!$acc& collapse(2)
      do i = 2, nbig
        do j = 2, nbod
          if (j.gt.i) then
            <calculation of minimal separation
            between each couple of bodies>
            if <min_sep>.le.<crit_dist> then
!$acc atomic capture
              nce = nce + 1
              this_nce = nce
!$acc end atomic
              ice(this_nce) = i
              jce(this_nce) = j
              ce(i) = 2
              ce(j) = 2
            end if
          end if
        end do
      end do
!$acc end parallel loop
\end{verbatim}
where \texttt{rcrit} is the critical radius (see Sect.~\ref{sss:main_loop}), \texttt{min\_sep} is the minimum separation of the bodies during the encounter and \texttt{crit\_dist} is the critical distance between the bodies that cause such an event to be flagged as a close encounter. \texttt{nce} is a scalar counting the global number of close encounters, while \texttt{ice}, \texttt{jce} and \texttt{ce} are three arrays storing the indexes of the two bodies undergoing a close encounter and a per-body flag. Loop 1 and 2 are straightforward and required no special information for the compiler to parallelize. Loop 3 is more complicated: we added the \texttt{collapse} keyword to flatten the nested loop and the \texttt{atomic} region to avoid data races on the scalar \texttt{nce} counter. The combination of collapsing the loops, which ensures that each CUDA core is managing a specific (i,j) couple of bodies, and the \texttt{atomic} region allows for the correct filling of the \texttt{ice}, \texttt{jce} and \texttt{ce} arrays.

As far as the parallelism granularity is concerned, we performed some experiments in setting explicitly the level of parallelism of either nested loops or, during development, of single unpacked subroutines. An example of the former was loop 3 of the \texttt{mce\_snif} routine, where we assigned the outer loop to gangs and the inner one to vectors. An example of the latter was the \texttt{mce\_min} routine, called in loop 3 of \texttt{mce\_snif}. This routine was tasked to perform a part of the calculation needed to find the minimum distance between two bodies, and had no internal loops. As such, we decorated it using, in one test \texttt{!\$acc routine vector}, and \texttt{!\$acc routine worker} in another. In all cases, the performances were either equal or worse than in the fully unpacked version of the code with no explicit assignment of the granularity. As such, we opted for this solution.

\begin{figure*}
\centering 
\includegraphics[width=0.48\textwidth]{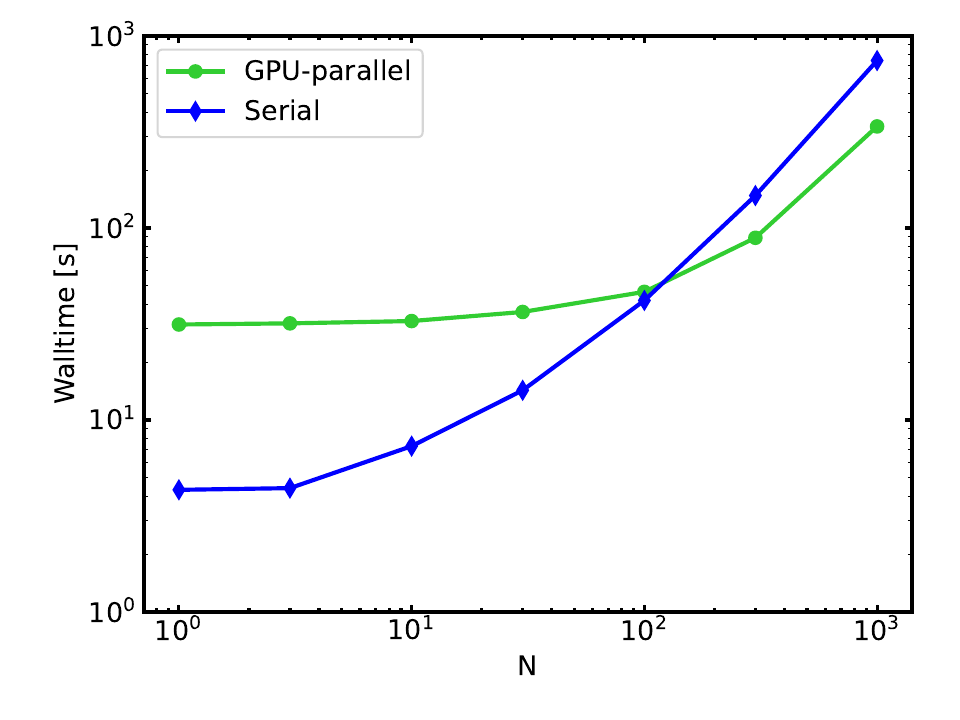}	
\includegraphics[width=0.48\textwidth]{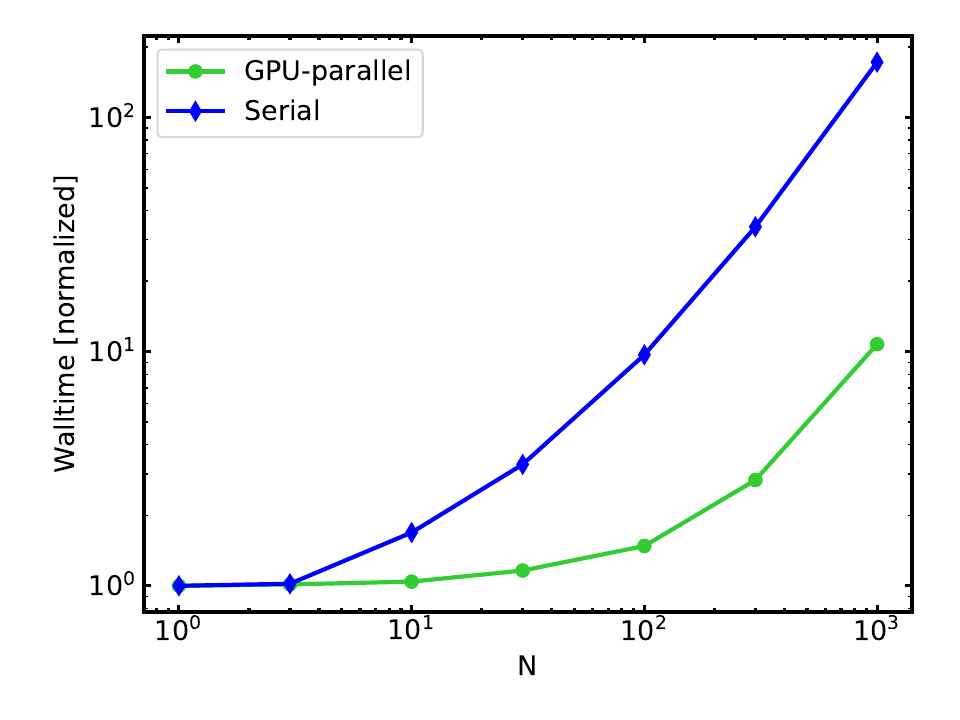}	
\caption{Left: the walltime $t_w$ required to run \mopal\, with an increasing amount of fully interacting bodies $N$ when executed in parallel on the GPU or serially on the CPU. The green line refers to the executable compiled by \texttt{nvfortran} of the Nvidia HPC SDK and decorated with OpenACC compiler directives. The blue line refers to the executable compiled by \texttt{ifx} Fortran compiler of the Intel OneAPI HPC  toolkit. Right: same, but times are normalized to the amount of time required to run each 1-body simulation.} 
\label{fig:performances}
\end{figure*}

\subsection{Data management}
\label{ss:mopal_data}

The data transfer between host and device is controlled by a data region encompassing the \texttt{mdt\_hy} subroutine. We defined the data transfer based on the following prescriptions:
\begin{enumerate}
\item scalar variables and constants that define the structure of the protoplanetary disk are copied from the host to the device only once at the first timestep and stored in the device memory;%beginning of the very first iteration;
\item local arrays to support the computation of the planet-disk interactions are created on the device at the first iteration and  stored in its memory;
\item the scalar variables containing the time, timestep and the number of bodies as well as the arrays of the mass, physical radius and critical radius values of the planetary bodies are copied from the host to the device. Masses and radii are updated only if the number of bodies change, since that might signal an impact (via the \texttt{dtflag} variable), while critical radii are always updated since they depend on the instantaneous distance between each planet and the central star;
\item local arrays storing information on close encounters (e.g., the list of bodies undergoing close encounters) are copied from the device to the host before calling \texttt{mdt\_hkce};
\item the position and velocity arrays and the timestep flag (that marks the direction of the integration) are copied one from the host to the device and back, at the beginning and at the end of the data region. If close encounters are detected by \texttt{mce\_snif}, positions and velocities are copied again twice at the beginning and at the end of \texttt{mdt\_hkce}.
%\item the position and velocity arrays and the timestep flag (that marks the direction of the integration) are copied twice from the host to the device (at the beginning of the data region and after the solution of the close encounters by \texttt{mdt\_hkce}), and twice from the device to the host (before calling \texttt{mdt\_hkce} and before exiting the data region);
\end{enumerate}

The data transfers between host and device therefore involve $7\times N$ IEEE-754 double precision floating point values (referred to solely as \textit{floats} hereafter) twice per iteration (three for positions, three for velocities and one for the critical radii for each body) at a minimum plus, potentially, $2\times N$ floats when the number of bodies change because of collisions in the previous timestep. Moreover, if a close encounter is detected, positions and velocities are exchanged again twice. While these are the most demanding arrays in terms of data transfer, they are not the only ones that must be exchanged between the host and the device. The identification of bodies undergoing close encounters is performed on the device, while the actual integration using the Bulirsch-Stoer algorithm and the impact identification are instead done on the host. As such, at each iteration, three integer arrays of length $N_{ce}$, which is the number of couples of bodies undergoing close encounters, containing the indices of the bodies undergoing close encounters and a flag related to the type of the close encounter must be transferred to the host. In practice, the code appear in this form:
\begin{verbatim}
!$acc parallel
      if (nce.gt.0) then
!$acc loop
        do j = 2, nbod
          <restore x and v to the state before
          the call of arxes_drift using x0 and 
          v0>
        end do
      end if
!$acc end parallel
!$acc update self(nce)
      if (nce.gt.0) then
!$acc update self(nce,tstart,ice(:nce),jce(:nce),
!$acc& ce(:nbod),x(:3,:nbod),v(:3,:nbod))
        <integrate close encounters on CPU>
!$acc update device(m(:nbod),x(:3,:nbod),v(:3,:nbod),
!$acc& hrec)
      end if
\end{verbatim}
where \texttt{nce} is $N_{ce}$, and \texttt{ice}, \texttt{jce} and \texttt{ce} are the arrays storing the indices of bodies undergoing close encounters.

%The calculation of the first part of the calculations are performed on the device. These calculations include advancing the Solar and interaction Hamiltonian for the fist half of the timestep, and the Keplerian Hamiltonian for the entirety of the timestep. After having checked for the close encounters (\texttt{mce\_snif}), an update of the positions and the velocities on the host is called, as it is the list of bodies flagged for close encounters. The Bulirsch-Stoer integration and the merging events (if they are present) are managed on the host, which then copies back to the device the new positions, velocities, masses and physical radii of the bodies. At this point, the last part of the integration, which include advancing the Solar and interaction Hamiltonian for the last half timestep, is performed.

\section{Test setup and results}
\label{sec:results}

\begin{figure*}
\centering 
\includegraphics[width=0.48\textwidth]{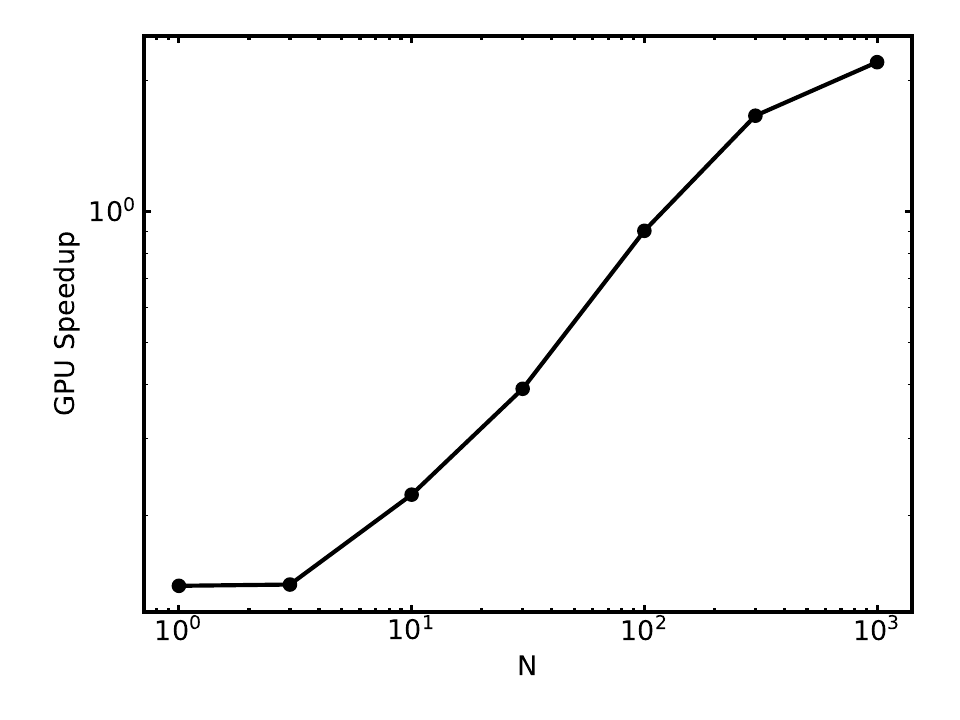}
\includegraphics[width=0.48\textwidth]{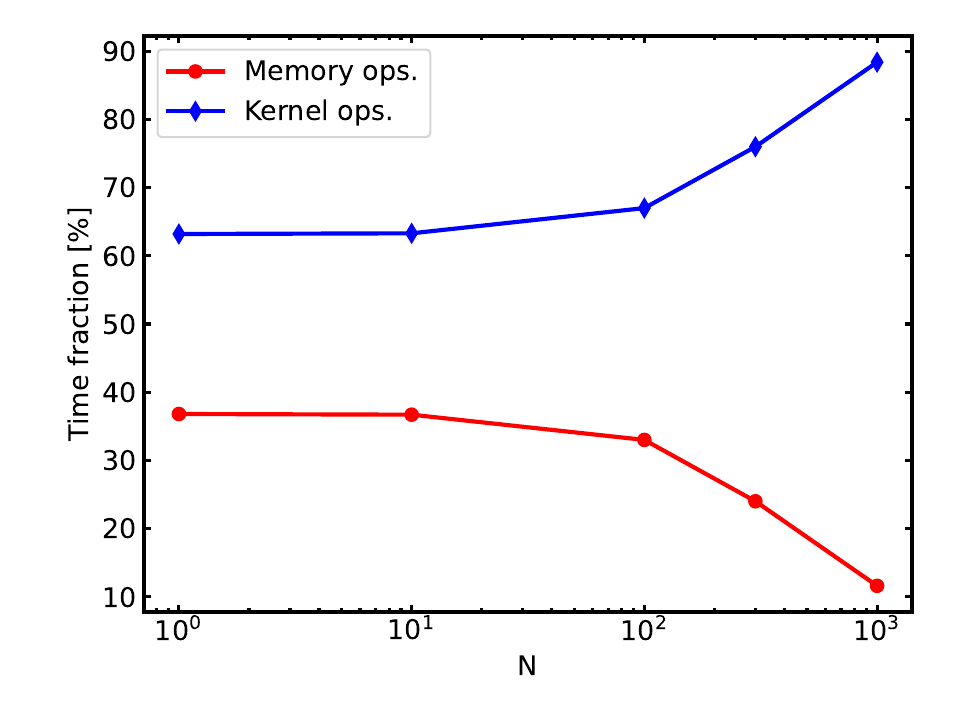} \\
\includegraphics[width=0.96\textwidth]{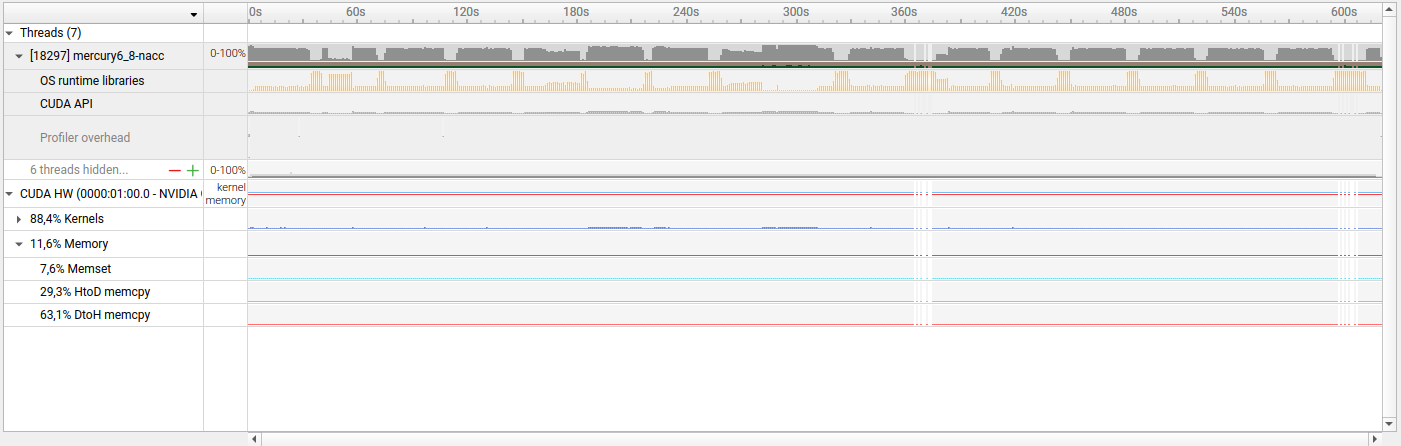}
\caption{Upper left: the speedup (on the left y-axis) and the relative speedup (on the right y-axis) of the GPU-parallel run of \mopal\, with respect to the serial run, as a function of the number of fully interactive bodies $N$. Upper right: the fraction of the total time spent in performing memory transfer operations (in red) and kernel operations (in blue), as derived by using the Nvidia \texttt{Nsight Systems} application. Lower panel: the timeline view in the \texttt{Nsight Systems} graphical user interface for a $N=1000$ case, showing how about 90\% of the GPU time is spent in calculations and 10\% in data transfers.} 
\label{fig:speedup}
\end{figure*}

We run with \mopal\,the same set of tests performed in Sect.~\ref{sss:marxes_prof} to study the performance of the code when running on GPU. We run the tests using the serial code and the GPU-based code on consumer-grade hardware, specifically an
%Both the serial and the parallel runs have been performed using a
%consumer laptop with a 
Intel Core i5-10200H CPU working at 2.40 GHz and a Nvidia GeForce RTX 3060 mobile (CUDA compute capability: 8.6\footnote{\url{https://developer.nvidia.com/cuda/gpus}})
%\footnote{At variance with the desktop version, the laptop version has a substantially lower working frequency, related to the thermal dissipation limitations of the machine. The power absorption is limited to 115 W versus the 170 W absorption of the desktop version.}) 
GPU with 3840 CUDA cores working in the 0.9-1.42 GHz range and capable of delivering up to 10.94 TFLOPS in floating point single precision operations\footnote{See technical data at \url{https://www.nvidia.com/en-us/geforce/laptops/compare/30-series/} and \url{https://www.techpowerup.com/gpu-specs/geforce-rtx-3060-mobile.c3757}}. As typical for consumer grade GPUs, the computing power when operating in double precision operations is significantly lower and decreases by a factor of 64 to 170 GFLOPS. 
%We note that computation using double precision floats on consumer GPUs is 64 times slower than single precision floats calculations, while for the CPUs the slowdown is only a factor of two. 
This fact substantially reduces the advantage of the GPU over the CPU, which is capable of delivering 154 GFLOPS in double precision\footnote{See technical data at \url{https://cdrdv2-public.intel.com/841556/APP-for-Intel-Core-Processors.pdf}} without the limiting factor of the data transfer between host and device.

The serial executable is compiled with the Fortran compiler of the Intel OneAPI HPC (\texttt{ifx} v.2025.2.1 20250806) toolkit while the GPU-based excutable is compiled with the Nvidia HPC SDK (\texttt{nvfortran} v.24.5-1, working on the \texttt{NVIDIA-SMI} driver v.580.95.05 and compiling GPU instructions in CUDA v.13.0). While the Nvidia HPC SDK is designed to fully take advantage of the Nvidia GPU capabilities, tests consistently showed that it significantly under-performs with respect to Intel OneAPI HPC toolkit when running on CPU. Both at hardware and simulation levels, therefore, the adopted setup is designed to stress test the performance of the GPU-based code under conditions that markedly favor the execution on CPU. Each test setup has been run five times and averaged over each set of five runs to minimize the impact of small performance fluctuations.

%when the number of floating points operations per second (FLOPs) are considered, even causing the final raw theoretical computation capacity of the GPU to be lower (0.108 TFLOPS\footnote{The official documentation on the consumer Ampere architecture can be found here: \url{https://www.nvidia.com/content/PDF/nvidia-ampere-ga-102-gpu-architecture-whitepaper-v2.1.pdf}. This, however, does not cover specifically the GA106 chipset of the Nvidia GeForce RTX 3060 (laptop), for which we have to calculate the the computing power indirectly from information on the Nvidia website.}) than the CPU (0.153 TFLOPS\footnote{\url{https://cdrdv2-public.intel.com/841556/APP-for-Intel-Core-Processors.pdf}}). Each configuration has been run five times to allow for small performance differences to be smoothed out.

\subsection{Execution times: GPU-parallel vs CPU-serial}
\label{ss:times}

The average runtime of the tests is reported in Figure \ref{fig:performances}, both as walltime $t_w$ (left) and as runtime normalized to the 1-body case $t_1$ (right). The GPU-parallel execution has a large disadvantage at low $N$ due to the large overhead in launching the kernel and copying the data for the first time from the host to the device, as well as the less performing code generated by the compiler for the parts executing on the CPU. Another source of disadvantage is that, at low $N$, the need for parallelism is low and the GPU is largely underutilized. These three causes manifest themselves in a substantially flat curve for the GPU wall time $t_w^{GPU}$ in the $N \in [1,100]$ range, with the execution time for the 100-bodies case lasting only a factor of 1.5 longer that the 1-body case. On the other hand, the CPU wall time $t_w^{CPU}$ of the serial run over the same $N$ range increases by a factor of nearly 10. Based on the number of CUDA cores present in the testing GPU, we reach full utilization (excluded the transfer times) at 62 bodies. For larger $N$, each CUDA core must work on more than one element of the $N\times N$ interaction matrix and this manifest itself in a change in the slope of the computation time curve. At $N=300$ bodies $t_w^{GPU}$ is 1.9 times larger than for $N=100$, and at $N=1000$ is 3.8 times larger that for $N=300$. The CPU curve also changes its slope, with $t_w^{CPU}$ increasing by a factor of 3.6 and 5.1 in the same $N$ interval.

The efficiency gain in using the GPU is more evident in the right panel of Figure \ref{fig:performances}. Normalizing to the 1-body case highlights the way GPU and CPU scale for this particular code and on this machine. The $N=1000$ bodies configuration in the GPU run of \mopal\, requires a $t_w$ which is $\sim11$ times more than the 1-body integration, while the serial run needs $\sim190$ more time. The advantage of the GPU parallel execution over the CPU serial execution can again be evaluated either in absolute terms, as the ratio between the wall time of the serial versus GPU-parallel execution $S_w=t_w^{CPU}/t_w^{GPU}$, commonly called speedup, or in terms of normalized times $S_1=t_1^{CPU}/t_1^{GPU}$, which we call here relative speedup. Both these quantities are reported in the upper left panel of Figure \ref{fig:speedup}. As it is possible to see, for $N \lesssim 100$ there is no advantage, in terms of wall time and in this very specific configuration, in running the \mopal\, on GPU, since $S_w<1$. In terms of $S_1$ however, the GPU-parallel execution is already 1.6 times faster at $N=10$. The GPU efficiency then continues to rise to break-even ($N=100$) and beyond, reaching $S_w=2.2$ and $S_1=16$ for $N=1000$. At $N=300$ it is possible to see an inflection point and the slope of the curve decreases.
The first cause we assumed for this behavior is the inefficient data transfer. However, a closer look at the time profile of the code, taken by means of the dedicated Nvidia \texttt{Nsight Systems}\footnote{\url{https://developer.nvidia.com/nsight-systems}} v.2024.1.1.59 toolkit, seems to disprove this possibility (see Figure \ref{fig:speedup}, upper right panel). In fact, the total wall time spent by \mopal\, when running on GPU in performing all memory transfer operations increases only from 2.89 s for $N=1$ to 4.47 s for $N=1000$, while the total execution time for the (profiled) runs are, respectively, 85 s and 620 s (Figure \ref{fig:speedup}, lower panel). The fraction of time spent in transfers thus decreases from 3.4\% to 0.7\% and it is expected to continue to decrease for larger simulations since it is $\propto N$. Similarly, the fraction of time spent in API calls is mostly constant. What really causes the loss of efficiency is the increasing amount of time the code spent in the \texttt{mdt\_hkce} routine, which is still executed on CPU.
%In fact, with this configuration, computation on GPU takes up less than 6\% of the wall time.
%This is possibly due to the inefficient data management due to the transfer between the host and the device, related to the choices of both keeping the close encounters integration on the CPU and the creation rate of simulation snapshots high.

\subsection{Energy conservation and numerical stability of the code}
\label{ss:conservation}

\begin{figure*}
\centering 
\includegraphics[width=0.48\textwidth]{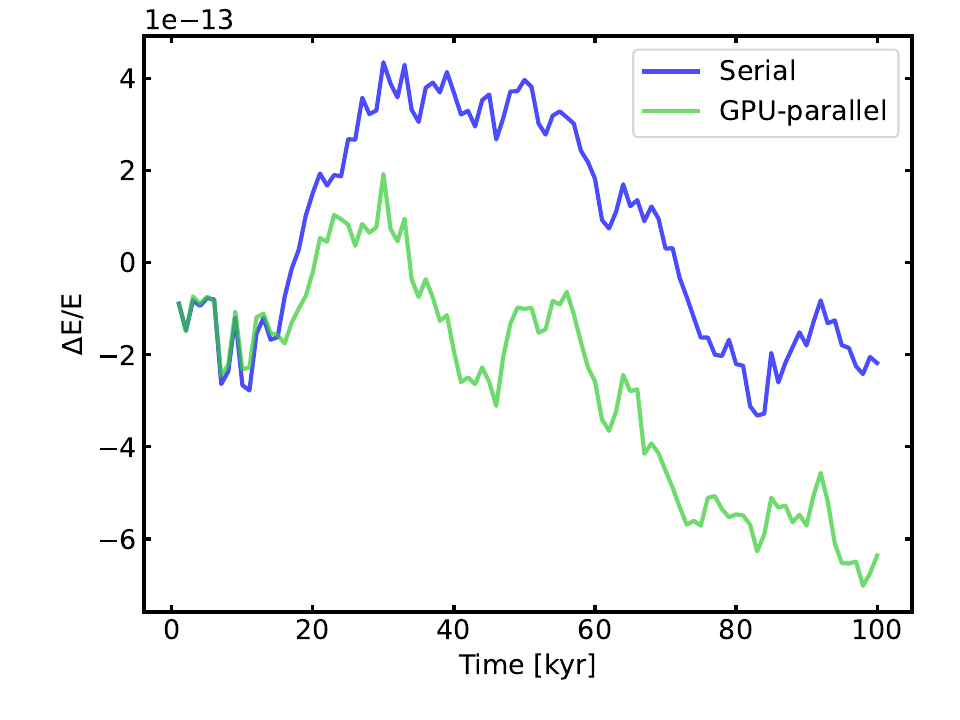}
\includegraphics[width=0.48\textwidth]{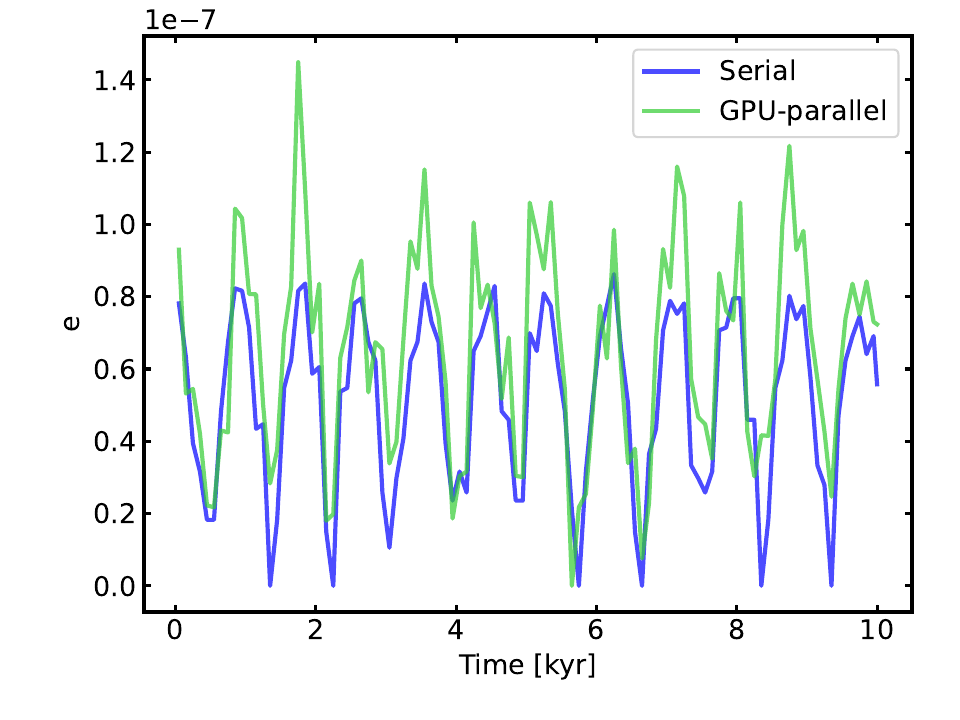}
\caption{Left: evolution of the relative energy oscillation over $10^5$ yr in a $N=1$ system (left). Right: evolution over $10^4$ yr of the eccentricity of the innermost planet in a $N=10$ system, which starts in a circular orbit. Blue and green curves refer, respectively, to serial (blue) and GPU-parallel (green) runs of \mopal.} 
\label{fig:conservation}
\end{figure*}

Due to the non-linear nature of n-body problems, we are interested in checking if our modifications impacted the ability of the code to conserve the total energy $E$ of the system and if there are significant deviations in the results of serial and GPU-parallel runs. In order to perform this analysis on \mopal, we ran three more tests: (i) a $N=1$ system with a Jupiter-sized planet in eccentric ($e=0.22$) orbit integrated for $10^5$ yr with no gas drag and gas gravity, to assess the pure numerical error introduced by the use of different hardware; (ii) a $N=10$ system integrated for $10^4$ yr with widely spaced Mars-sized objects in circular orbits, to check how much the numerical error on small planet-planet interactions affect the long term evolution of their orbital parameters and (iii) a $N=100$ densely packed system integrated for $10^4$ yr, to compare the final distributions of $a$ and eccentricies $e$ produced by a mix of long distance interactions and close encounters.

The results for tests (i) and (ii) are reported, respectively, in the left and right panels of Figure \ref{fig:conservation}. Quantities are sampled every $10^3$ yrs in both cases. As it is possible to see in the left panel, energy is conserved within a few $10^{-13}$ both in the serial and in the GPU-parallel run, and its evolution over time is substantially similar. This level of precision is expected for a two-body system where only the keplerian component of the Hamiltonian is non-zero: similar results are obtained running the original \marxes\, code with the \texttt{WHFAST} routine, and are comparable to those reported by \citet{rein2015}. Peak $|\Delta E/E|$ is slightly higher in the GPU-parallel run, but it is very likely due to the random evolution of the energy error, since both curves represent a single run. The impact of numerical error on the orbital parameters of the bodies is equally limited. In the right panel of Figure \ref{fig:conservation} we show, as an example, the eccentricity evolution of the innermost planet of case (ii) starting from an initial circular orbit. Eccentricity variations are well reproduced and are of the same order of magnitude for the serial and the GPU-parallel case. While not shown, the same applies also to the semi-major axis evolution.

\begin{figure}
\centering 
\includegraphics[width=0.48\textwidth]{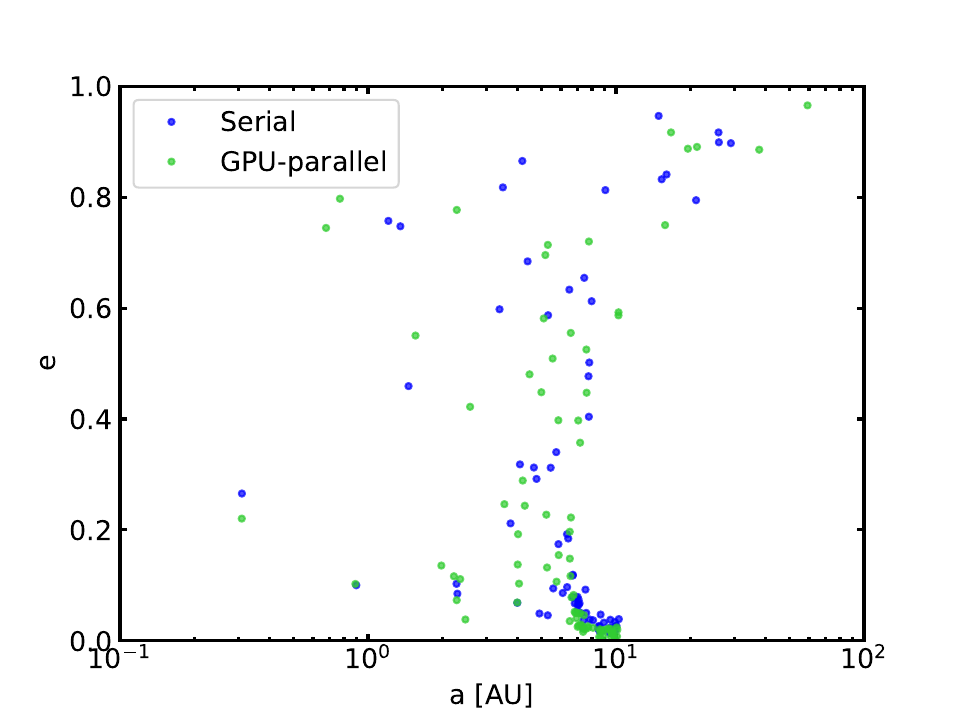}
\caption{Configuration in the semi-major axis-eccentricity plane of a 100-body system after 10 kyr for serial (blue) and GPU-parallel (green) runs.} 
\label{fig:scatter_results}
\end{figure}

Finally, in Figure \ref{fig:scatter_results} we show the final dynamical state in the a-e plane of the $N=100$ system after a 10 kyr long integration. As it is possible to see, the results are comparable: in both the serial and the GPU-parallel run,  bodies up to $\sim$6 AU are dynamically excited on high eccentricity orbits and the boundary of the parameter space filled by them is well defined. The median values of $a$ in the serial and parallel runs are, respectively, 7.07 and 7.04 AU, while the median values of $e$ are, respectively, 0.1002 and 0.1113.

\section{Discussion}
\label{sec:discussion}

As mentioned in the Introduction, OpenACC is not the only parallelization framework in existence. Its choice for this work was driven by considerations of portability, accessibility and serviceability. In order to understand our choice, a good starting point is to consider that these three concepts are not completely independent from one another, since one of the ways to achieve a high level of portability is to ensure robust accessibility from the community and short development cycles of specialized versions of the code, each capable of extracting the best performances from each platform \citep[see][for a thorough discussion on the subject]{pennycook2019}. This reasoning can be further expanded by considering the development and optimization time, alongside the execution time, when accounting for the global performance of a given code. While this point might not be relevant from a purely HPC perspective, it is integral to the scientific cost-benefit evaluation of parallelizing and porting on GPU any already existing model. It is also worth mentioning that while high-level, directive-based frameworks does not necessarily score high in the portability performance metric delineated by Pennycook and collaborators \citep[][]{malenza2024}, it has been shown that they perform increasingly better on the type of large numerical problems that \mopal\, has been developed to tackle, due to higher occupancy fractions on the GPU \citep[][]{malenza2025}. As such, considering both the extended interpretation of portability described above and the small expected loss in performance with respect to a lower level porting, OpenACC offered the best compromise for the case presented here.

Another choice that warrants discussion is why we decided to port anew to GPU a code when other GPU-accelerated n-body integrators exist. The answer lies in our qualitative cost-benefit analysis under the same theoretical framework presented in \citet{pennycook2019}: despite being already GPU-ported, codes like \texttt{GENGA II} and \texttt{GLISSE} do not include the same physics recipes as \marxes. To integrate the same physical modeling of the protoplanetary disk in an already optimized code in CUDA would have resulted in a larger development time requirement and an overall lower final portability of the code.

\section{Conclusions}
\label{sec:conclusion}

In this paper we presented \mopal\,, the first GPU-accelerated version of the \marxes\, n-body code for planetary formation simulations. \mopal\, is ported to GPU computing with OpenACC to ensure cross-platform compatibility and minimize the need for deep code restructuring.  
%, thus maximizes code accessibility to the research community. -> eviterei di inserirlo perché al momento non rilasciamo il codice.
Tests performed on consumer-grade hardware for 1-1000 fully interacting bodies embedded in their native protoplanetary disk demonstrate that, despite the suboptimal test configuration we adopted, the GPU execution is up to 2.2 times faster than the serial execution and that the wall time scales appropriately with the number of CUDA cores in the GPU. We checked the ability of the code to conserve energy over a $10^5$ yr baseline, confirming its ability to carry out accurate planetary formation simulations that require $\mathcal{O}(10^6)$ yr of integration time. As \marxes\,, \mopal\, is available to the community through collaborations with the developing team.

\section*{Acknowledgements}
The authors thank the anonymous referees for their valuable comments. This work is supported by the Fondazione ICSC, Spoke 3 “Astrophysics and Cosmos Observations'', National Recovery and Resilience Plan (Piano Nazionale di Ripresa e Resilienza, PNRR) Project ID CN\_00000013 “Italian Research Center on High-Performance Computing, Big Data and Quantum Computing'' funded by MUR Missione 4 Componente 2 Investimento 1.4: Potenziamento strutture di ricerca e creazione di “campioni nazionali di R\&S (M4C2-19)'' - Next Generation EU (NGEU). The authors acknowledge support from the ASI-INAF grant no. 2021-5-HH.0 plus addenda no. 2021-5-HH.1-2022 and 2021-5-HH.2-2024, the INAF Main Stream project “Ariel and the astrochemical link between circumstellar discs and planets” (CUP: C54I19000700005), the COST Action CA22133 PLANETS, and the European Research Council via the Horizon 2020 Framework Programme ERC Synergy “ECOGAL” Project GA-855130. The authors also acknowledge the support of Amazon Web Services in the form of time allocation on their AWS EC2 infrastructure to the INAF-ICT2018 pilot study on OpenACC. The authors wish to thank Martina Vicinanza for her support to the INAF-ICT2018 pilot study as well as Francesco Reale and Mirko Riazzoli for their support in managing the Genesis computational cluster at INAF, which supported the early development phase of {\sc Mercury-Opal}.
\appendix

%\section{Appendix title 1}

%\section{Appendix title 2}

\bibliographystyle{elsarticle-harv} 
\bibliography{example}



%% For example:

%% \bibitem[Aladro et al.(2015)]{Aladro15} Aladro, R., Martín, S., Riquelme, D., et al. 2015, \aas, 579, A101

%%\end{thebibliography}

\end{document}